\documentclass[aps,pra,superscriptaddress,twocolumn,showpacs,10pt]{revtex4}

\usepackage{amsmath,amsthm,amssymb}
\usepackage{graphicx}
\usepackage{dcolumn}
\usepackage{bm}
\usepackage{color} %for change tracking

\usepackage{graphicx,graphics}
\usepackage{dcolumn}				% Align table columns on decimal point
\usepackage{amssymb}
\usepackage{bm,pstricks}					% bold math

\newcommand{\beqa}{\begin{eqnarray}}
\newcommand{\eeqa}{\end{eqnarray}}
\newcommand{\beq}{\begin{equation}}
\newcommand{\eeq}{\end{equation}}

\newcommand{\ket}[1]{| #1 \rangle}

\begin{document}

\title{Sequential measurement-based quantum computing with memories}

\author{Augusto J. Roncaglia}
\affiliation{ICFO-Institut de Ci\`{e}ncies Fot\`{o}niques, Mediterranean Technology Park, 08860 Castelldefels (Barcelona), Spain}

\author{Leandro Aolita}
\affiliation{ICFO-Institut de Ci\`{e}ncies Fot\`{o}niques, Mediterranean Technology Park, 08860 Castelldefels (Barcelona), Spain}

\author{Alessandro Ferraro}
\affiliation{Grup d'\`{O}ptica, Universitat Aut\`{o}noma de Barcelona, E-08193 Bellaterra (Barcelona), Spain}
\affiliation{Department of Physics and Astronomy, University College London, Gower Street, London WC1E 6BT, United Kingdom}

\author{Antonio Ac\'in}
\affiliation{ICFO-Institut de Ci\`{e}ncies Fot\`{o}niques, Mediterranean Technology Park, 08860 Castelldefels (Barcelona), Spain}
\affiliation{ICREA-Instituci\'o Catalana de Recerca i Estudis Avan\c cats, Lluis Companys 23, 08010 Barcelona, Spain}

%%%%%%%%%%%%%%%%%%%%%%%%%%%%%%%%%%%%%%%%%%%%%%%%%%%%%%%%
%%%%%%%%%%%%%%%%%%%%%%%%%%%%%%%%%%%%%%%%%%%%%%%%%%%%%%%%
\begin{abstract}
We introduce a general scheme for sequential one-way quantum computation where static systems with long-living quantum coherence
(memories) interact with moving systems that may possess very short coherence times. Both the generation of the cluster state
needed for the computation and its consumption by measurements are carried out simultaneously. As a consequence, effective
clusters of one spatial dimension fewer than in the standard approach are sufficient for computation.
In particular, universal computation requires only a one-dimensional array of memories.
The scheme applies to discrete-variable systems of any dimension as well as to continuous-variable ones,
and both are treated equivalently under the light of local complementation of graphs. 
In this way our formalism introduces a general framework that encompasses and generalizes in a 
unified manner some previous system-dependent proposals.
The procedure is intrinsically well-suited for implementations with atom-photon interfaces.
\end{abstract}


\maketitle
%%%%%%%%%%%%%%%%%%%%%%%%%%%%%%%%%%%%%%%%%%%%%%%%%%%%%%%%%%%%%%%%%%%%%%%%%%%%%%%%%%%%%%%%%%%%%%%%%%%%%%%%%%%%%
%%%%%%%%%%%%%%%%%%%%%%%%%%%%%%%%%%%%%%%%%%%%%%%%%%%%%%%%%%%%%%%%%%%%%%%%%%%%%%%%%%%%%%%%%%%%%%%%%%%%%%%%%%%%%

\section{Introduction}
Arguably the most ambitious enterprise of
quantum information science is the realization of a universal
quantum computer. That is, a machine able to efficiently process
quantum information in any desired fashion. Historically, this
quest was first pursued inspired by the classical notion of an
array of logical gates --a circuit--  that sequentially act on a
physical system encoding the logical state of the computer
\cite{NielsenChuang}. In this approach, the {\it circuit model},
all gates are decomposed into one-body (local) and two-body
(entangling) unitary operations, and the output is read out by
measuring the system's final state. A major advantage of the
circuit model is that it requires the generation and
coherent preservation of many-body entanglement only on a number
of system constituents linear with the logical inputs. In what
follows, we refer to this as Property $P1$. On the other hand,
the sequence of operations must be adapted to each given
computation, which represents a considerable obstacle from an
experimental viewpoint. In particular, entangling operations must
be actively applied {\it on-line} --during the course of the
computation-- while simultaneously preserving the system's quantum
coherences.

\par Recently, a conceptually different paradigm has been put forward:
measurement-based quantum computation (MBQC) \cite{1WQC}.
There, the computation is performed by adaptive local measurements
on certain quantum lattices in universal many-body-entangled
resource states. The clear advantage is that no demanding on-line
operations are required throughout the computation: only
measurement bases must be (locally) adapted, whereas the
generation of the resource state is independent of the given
computation. More specifically, information processing
proceeds via local measurements only (Property $P2$); and the
architecture of the interactions needed for the resource-state
generation is independent of the given computation (Property
$P3$). The price to pay however is that universality is attained
only with two-dimensional lattices, implying that the number of
system components typically scales quadratically with the logical
inputs. To make things worse, such many-body entangled states seem
not to be trivial to find in natural systems \cite{HamsNoGo}. All in all, and despite the
success in singling them out as ground states of specific, relatively simple
Hamiltonians \cite{1WQC, Hamiltonians}, their generation or
coherent preservation still remain a challenge.

\begin{figure}[b]
\begin{center}
\includegraphics[width=0.65\linewidth]{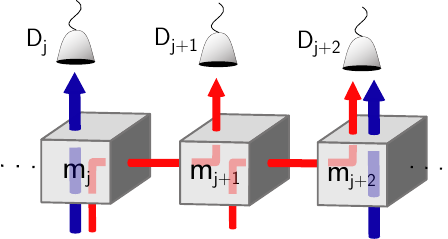}
\caption{Pictorial sketch of our sequential model for quantum computing.
Long-lived quantum systems $m_j$ --{\it quantum memories}-- iteratively
interact at fixed interfaces with moving registers --{\it flying registers} (curves).
Registers do not need to be temporally synchronized and can also mediate the interaction between memories (thin curves).
At each step, a graph state is generated and part of it immediately consumed by measurements at detectors $D_j$. Several
physical platforms are envisaged for implementations in both discrete and continuous variable systems (see text).
\label{exp}}
\end{center}
\end{figure}

In this work we introduce a hybrid model of quantum
computation that combines the main practical advantages of the
circuit and MBQC models, and satisfies concurrently all the three
properties $P1-P3$ listed above. The key idea of the construction
is to apply the concept of local complementation of
graphs~\cite{LCgraphs,BOUCHET, graph_review, Zhang} to the most studied MBQC
architecture, namely the one-way quantum computer (1WQC) \cite{1WQC}. 
The resulting model preserves the advantages of
1WQC, while it relaxes the main obstacle toward its
implementation. The procedure is completely general and can be
applied both to discrete-variable (DV) systems of {\it any}
dimension \cite{1WQC,highD} and continuous variables (CVs)
\cite{1WMenicucci}.

Our proposal consists of an array of static, long-lived
quantum registers --{\it quantum memories} \cite{comment}-- that
sequentially interact with moving, short-lived quantum registers
--{\it flying registers}-- via a fixed interaction (the only
entangling operation required), see Fig. \ref{exp}. At each step
of the sequence, a graph state \cite{graph_review} (defined below)
among memories and flying registers is generated and, immediately
afterwards, the flying sub-graph is consumed by adaptive local
measurements. The architecture of the system operations stays the
same throughout and only the measurement bases must be adapted at
each step, preserving the main advantages of the 1WQC.
However, in contrast to the latter, here quantum information does
not flow from a physical site to another but 
in each step is stored in the memories. These are the only
quantum systems required to possess full coherence robustness
along the computation. Thus, a number of memories equal to the
logical inputs suffices for universal computation.

The present approach is inspired and in turn sheds light
on other recent proposals for hybrid quantum  computing. For the case
of two-dimensional systems, our formalism recovers the
ancilla-driven model \cite{Anders09}. This model, originally
derived within a different framework, also satisfies
properties $P1-P3$. The main advantage here is that, while the
relation between the ancilla-driven approach and 1WQC is not
straightforward~\cite{Anders09,Kashefi09}, our derivation
immediately enlightens this equivalence. This enables the direct
application of topological fault-tolerant techniques already developed for
standard 1WQC to our approach~\cite{FaultTolerant}. For the CV case, in turn, our
formalism recovers some aspects of the experimental proposal of
Ref.~\cite{Menicucci10}. There, an implementation of the CV 1WQC
model was put forward. Nevertheless, this proposal does not
satisfy property ($P1$) and requires pulse synchronization, long
fiber loops, and non-linear photon-photon interactions, which
impose serious technical obstacles. Our proposal alleviates these
experimental requirements and can be realized using experimentally
accessible quantum interfaces~\cite{Polzik}. 
In addition, the formalism  presented here sets a  
framework within which the proposals of Refs. \cite{Anders09,Kashefi09,Menicucci10}
can be understood, and extended to DV systems of arbitrary dimension, in a unified way.

The paper is organized as follows. In Sec. \ref{Sec-graphstates} we review some basic properties of graph states 
and their local complementation, for DV and CV systems. There, in particular, we also derive an equivalence between
the SWAP operation and successive local complementations for the case of vertices with a single neighbor.
In Sec. \ref{Sec-Sequential} we describe our unified picture for sequential MBQC. Physical setups 
for the implementations of our scheme are listed in Sec. \ref{Imp}. Finally, we present our conclusions in Sec. \ref{Sec-Conclusions}.

\section{Graph states} \label{Sec-graphstates}
As said, we focus throughout on 1WQC.
Its resource states are the celebrated cluster states \cite{1WQC, 1WMenicucci, cluster, highD}, which are graph states
\cite{graph_review} corresponding to square graphs. Before introducing graph states, let us recall the definition of
mathematical graphs: Consider the union $G_{(\mathcal{V},\mathcal{E})}\equiv\{\mathcal{V},\mathcal{E}\}$ of a set $\mathcal{V}$
of $|\mathcal{V}|=N$ vertices and a set $\mathcal{E}$ of edges $\{j,k\}$ with $j,k\in\mathcal{V}$. This can be univocally
represented by an $N\times N$ {\it adjacency matrix} $\Gamma$, of elements $\Gamma_{jk}\in\mathbb{R}_{\neq0}$ if $\{j,k\}\in \mathcal{E}$, and
$\Gamma_{jk}=0$ otherwise. If $\Gamma_{jk}=1\ \forall\ \{j,k\}\in \mathcal{E}$ we call $G_{(\mathcal{V},\mathcal{E})}$ an
{\it unweighted graph} --or simply a {\it graph}-- and denote it by $G$ for short; otherwise we write
explicitly  $G_{(\Gamma)}$ and say that the graph is {\it weighted}, understanding that each edge $\{j,k\}$ has an associated
weight $\Gamma_{jk}$.

\par In both the DV and CV regimes, a graph state $\ket{G}$ can be defined as follows: ({\it i}) To each  vertex $j\in \mathcal{V}$
associate a  quantum system initially in a state $\ket{+}_j$, given by the uniform superposition of all computational states.
For example, in the case of qubits $\ket {+}_j=\hat H_j\ket{0}_j\equiv\frac{1}{\sqrt{2}}(\ket{0}_j+\ket{1}_j)$ is the eigenstate of the $\hat{X}_j$
Pauli operator with positive eigenvalue,  $\hat H$ is the Hadamard transform, and the computational states $\ket{0}_j$ and $\ket{1}_j$
are respectively the eigenstates with positive and negative eigenvalues of the $\hat{Z}_j$ Pauli operator.
For CV systems in turn, $\ket {+}_j=\ket{0}_{p_j}=\hat{F}_j\ket{0}_{q_j}\equiv\frac{1}{\sqrt{2\pi}}\int_{\mathbb{R}}du\ket{u}_{q_j}$
represents the zero-momentum eigenstate of the momentum quadrature operator $\hat{p}_j$. In the last, the computational states
$\ket{u}_{q_j}$ are the eigenstates of the position quadrature operator $\hat{q}_j$, with  $[\hat{q}_j,\hat{p}_j]=i\mathbf{1}$
(we take $\hbar\equiv1$ throughout), and $\hat{F}_j$ is the Fourier transform on the $j$-th quantum mode (qumode). ({\it ii})
For every edge $\{j,k\}\in\mathcal{E}$ apply a (maximally entangling) controlled-$Z$ gate $\hat{CZ}_{jk}$ to registers $j$ and $k$.
For the case of qubits, it is
$\hat{CZ}_{jk}=e^{i\frac{\pi}{4} (\mathbf{1}-\hat{Z}_j)\otimes(\mathbf{1}-\hat{Z}_k)}$.
This definition also extends directly to graph states on $d$-dimensional systems (qudits) \cite{highD}. For qumodes in turn,
it is $\hat{CZ}_{jk}\equiv e^{i\hat{q}_{j}\otimes\hat{q}_k}$. In all cases it is legitimate to write
$\ket{G}=\prod_{\{j,k\} \in \mathcal{E}} \hat{CZ}_{jk}\bigotimes_{j\in\mathcal{V}} \ket{+}_j$.
Finally, weighted graph states are obtained by introducing multiplicative factors $\Gamma_{ij}$ in the
exponents of the entangling operations.

\begin{figure}[t]
\begin{center}
\includegraphics[width=1\linewidth]{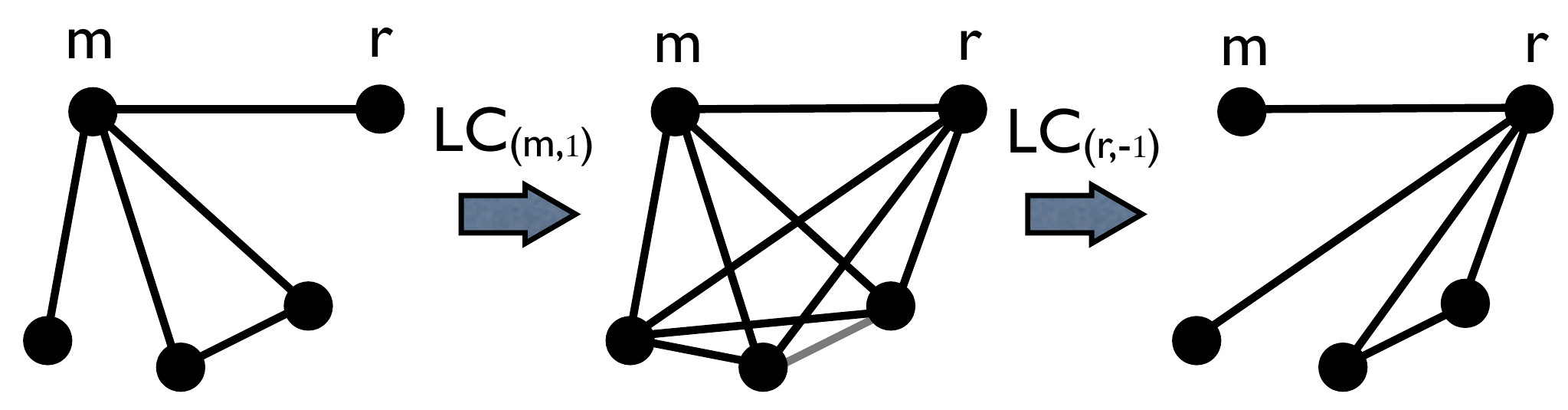}
\caption{Two sequential local complementations: first at vertex $m$ with weight 1 (left) and then (on the resulting graph)
at vertex $r$ with weight -1 (right). Black and gray lines correspond to graph edges with weight 1 and 2, respectively.
When the initial graph is unweighted and  $m$ is the only neighbor of $r$, such composition of local complementations
is always equivalent to a SWAP operation between vertices $m$ and $r$. \label{LC_SWAP}}
\end{center}
\end{figure}

\subsection{Local complementations of graphs and the SWAP operation} 
\par A key family of graph transformations is that of local
complementations \cite{BOUCHET}. The  definition that we use follows closely (and is a simplified version of)
the one in Ref. \cite{Zhang}, more naturally-motivated for
weighted graphs. We define the local complementation $LC_{(l,\delta)}$ of $G_{(\Gamma)}$, at node $l\in \mathcal{V}$, and of
real weight $\delta$, as the map  $G_{(\Gamma)}\rightarrow G_{(\Gamma'_{(l,\delta)})}$. %\equiv G'_{(l,\delta)}.
Here,  ${\Gamma'_{(l,\delta)}}_{jk}=\Gamma_{jk}+\delta$, if $\{j,k\}\in\mathcal{N}_{l}$, and
${\Gamma'_{(l,\delta)}}_{jk}=\Gamma_{jk}$ otherwise. The set $\mathcal{N}_{l}$ is the neighborhood of $l$
(in $G_{(\Gamma)}$), composed of all vertices $n\in \mathcal V$ connected to $l$ by some edge $\{l,n\}\in \mathcal E$.
In Fig. \ref{LC_SWAP} we show an example of two sequential local complementations. An immediate but very powerful
observation follows for all unweighted graphs where one of their vertices --say $r$-- has only one neighbor --say $m$--:
The operation $LC_{(m,1)}$ followed by $LC_{(r,-1)}$ is always equivalent to the exchange (SWAP) $m\leftrightarrow r$.

\par Now, every mathematical transformation $LC_{(l,\delta)}$ has a unitary representation $\hat{U}_{LC_{(l,\delta)}}$
(mapping  $\ket{G_{(\Gamma)}}$ into $\ket{G_{(\Gamma'_{(l,\delta)})}}$) which turns out to be {\it local} for the aforementioned SWAP case.
Therefore, if we consider the local complementations in Fig. \ref{LC_SWAP}, the total transformation, for the qubit case, can be
expressed in terms of {\it two simple local} operations: 
\beq
\label{qubit}
\hat{U}_{T}=\hat{U}_r\otimes\hat{U}_m=\hat H_r \otimes \hat H_m.
\eeq
For qumodes, the above transformation is implemented similarly as 
\beq
\label{qumode}
\hat{U}_{T}=\hat{U}_r\otimes\hat{U}_m=\hat F_r^\dagger \otimes \hat F_m.
\eeq
All these considerations hold true not only for qubits and qumodes but also
extend straightforwardly to qudit graph states \cite{highD}, where the role of the Hadamard and Fourier transform gates is 
played by the discrete Fourier transform (recall that the Hadamard is the discrete Fourier transform for dimension 2). 
In what follows we give  explicit proofs just for  the qubit and qumode cases, {\it i.e.} identities \eqref{qubit} and \eqref{qumode}.

\par As we said before, every mathematical transformation $LC_{(l,\delta)}$ has a unitary representation $\hat{U}_{LC_{(l,\delta)}}$
mapping  $\ket{G_{(\Gamma)}}$ into $\ket{G_{(\Gamma'_{(l,\delta)})}}=\hat{U}_{LC_{(l,\delta)}}\ket{G_{(\Gamma)}}$.
Remarkably, this unitary operation turns out to be {\it local}. In the case of graph-states of qubits for instance, it is 
$\hat{U}_{LC_{(j,\pm1)}}=e^{\mp i\frac{\pi}{4}\hat{X}_j}\bigotimes_{l\in\mathcal{N}_j}e^{\pm i\frac{\pi}{4}\hat{Z}_l}$
\cite{graph_review}, which belongs to the group of local Clifford operations. Thus, the considered succession of local 
complementations is represented by
\beqa
\label{totalqubit}
\nonumber\hat{U}_{T}&=&\hat{U}_{LC_{(r,-1)}}\hat{U}_{LC_{(m,1)}}\\
\nonumber&=&\Bigg(e^{i\frac{\pi}{4}\hat{X}_r}\bigotimes_{l\in\mathcal{N}'_r}e^{-i\frac{\pi}{4}\hat{Z}_l}\Bigg)\left( e^{-i\frac{\pi}{4}\hat{X}_m}\bigotimes_{l\in\mathcal{N}_m}e^{i\frac{\pi}{4}\hat{Z}_l}\right) \nonumber \\
&=&\left(\hat H_r \otimes \hat H_m\right)\left(e^{-i\frac{\pi}{4}\hat X_r}\otimes e^{i\frac{\pi}{4}\hat Z_m} \right). 
\eeqa
Notice that the primed neighborhood $\mathcal{N}'_r$ refers to the neighborhood of $r$ in the graph $G_{(\Gamma'_{(m,1)})}$ resulting 
from the first complementation. In addition, the operator
$e^{-i\frac{\pi}{4}\hat X_r} e^{i\frac{\pi}{4}\hat Z_m}$ \emph{stabilizes} all unweighted graphs having a vertex $r$ with a single neighbor $m$.
This can be easily checked by noticing that 
$e^{-i\frac{\pi}{4}\hat X_r} e^{i\frac{\pi}{4}\hat Z_m} \hat{CZ}_{rm}=\hat{CZ}_{rm}e^{i\frac{\pi}{4}\hat Z_m} e^{-i\frac{\pi}{4}\hat X_r \hat Z_m}$,
and that the register always starts in the state $\ket +$. Therefore, restricted to the states under consideration, 
the total transformation \eqref{totalqubit} is identical to \eqref{qubit}.

\par For qumodes in turn, the corresponding unitary transformation is
$\hat{U}_{LC_{(j,\pm1)}}=e^{\pm\frac{i}{2}\hat{p}^2_j}\bigotimes_{l\in\mathcal{N}_j}e^{\mp\frac{i}{2}\hat{q}^2_l}$ \cite{Zhang}, 
which belongs to the group of local Gaussian operations. 
The total composition is now
\beqa
\label{totalqumode}
\nonumber\hat{U}_{T} %&\equiv&\hat{U}_{LC_{(r,-1)}}\hat{U}_{LC_{(m,1)}}\nonumber \\ 
&=& e^{-\frac{i}{2}\hat{p}^2_r}e^{-\frac{i}{2}\hat{q}^2_r}\otimes e^{\frac{i}{2}\hat{q}^2_m}e^{\frac{i}{2}\hat{p}^2_m} \nonumber \\
&=&\left(\hat F_r^\dagger \otimes \hat F_m\right) \left( e^{\frac{i}{2}\hat{p}^2_r}\otimes e^{-\frac{i}{2}\hat{q}^2_m}\right).
\eeqa
As in the previous case, the operator at the right of the Fourier transforms stabilizes the graph states in question, since
$e^{\frac{i}{2}\hat{p}^2_r}e^{-\frac{i}{2}\hat{q}^2_m}\hat{CZ}_{rm}=\hat{CZ}_{rm}e^{\frac{i}{2}\hat{p}^2_r} e^{i\hat{p}_r \hat{q}_m}$, and the
register is always in the state $\ket{p=0}$. 
Thus the total transformation \eqref{totalqumode} is identical to \eqref{qumode}.

\begin{figure}[t]
\begin{center}
\includegraphics[width=0.8\linewidth]{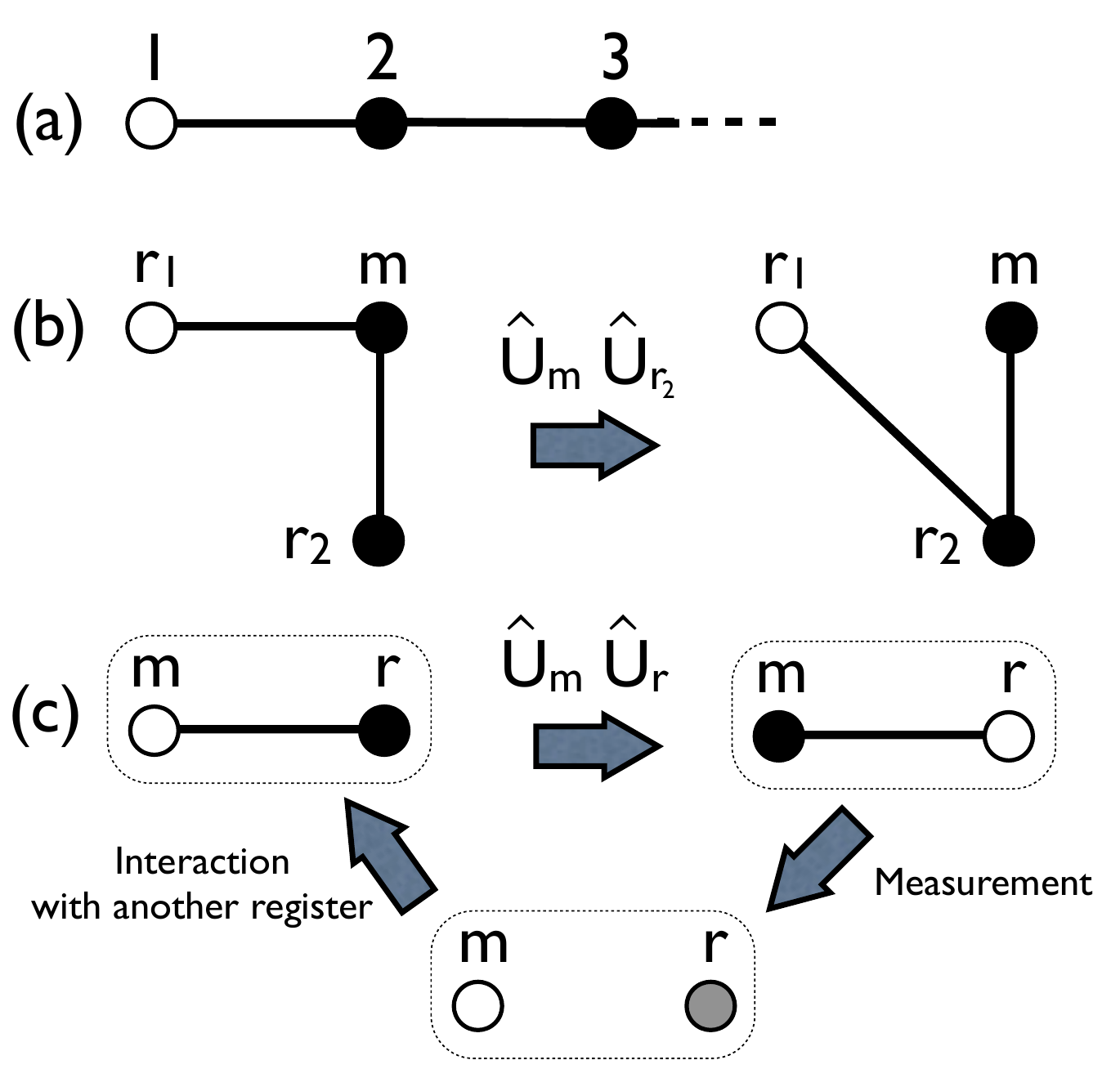}
\caption{1WQC along a quantum wire. The black circles represent systems prepared in
$\ket +$ and the black segments entangling gates. The node to be measured is depicted in white, and the gray circles represent
 measured nodes.
$(a)$ In the original scheme measurements are performed from left to right along a long linear cluster-state prepared in advance.
$(b)$ In the considered scenario a static quantum memory $m$ interacts sequentially with the flying register $r_{1}$,
then with $r_2$, etc., and a star-graph state with $m$ in the center is created (left). Local complementations on $m$
and $r_{2}$ swap the positions of $m$ and $r_2$ in the associated graph, thus transforming it into a linear graph with $m$
at the rightmost end (right).
$(c)$ Since all the operations (measurements and local complementations) involving different flying registers commute, the same
computation can be implemented sequentially by iteration of the cycle shown.
Notice further that there is no need to  actively apply $\hat{U}_r$, since it can always be absorbed in the measurement
by means of a passive basis-redefinition.
\label{cluster1d}}
\end{center}
\end{figure}

\section{Sequential MBQC}\label{Sec-Sequential}
\par Before introducing our scheme, let us briefly recall the conventional approach for the simplest building block of the 1WQC,
the 1D cluster state, also called quantum wire [see Fig. \ref{cluster1d}(a)]: Once the cluster is available the computation
proceeds by adaptive local measurements along it, from left to right in the figure. The measurement-basis choices and the
length of the wire depend on the particular computation to perform. Information flows from left to right until all but the
particle at the rightmost end is measured. The latter encodes the output, and any single-particle unitary operation can
be implemented in such a way.

\par  As said, the scenario we are interested in involves instead a static long-lived memory $m$ that sequentially interacts
with  flying registers $r_i$. The latter are not required to possess robust coherence properties, as they are measured
immediately after the interaction. All systems are initially in state $\ket +$ and each interaction drives a $\hat{CZ}$ gate,
so that a graph state among the memory and the passing registers is created after each interaction. Our aim is to reproduce
the conventional 1WQC approach but with the following variations: only the smallest necessary pieces of the
resource state are created at each step of the sequence; and information is stored and processed always in the same system, the memory.
This would be impossible if the flying registers were simply measured as they come out from the interaction region. The reason
is that the considered sequence creates a graph state which is not linear but star-like (with $m$ at the center),
thus not useful for arbitrary single-particle unitary operations.

\par A natural way to circumvent this issue is to swap the positions (in the underlying graph) of $m$ and $r_{i-1}$ before the $i$-th
interaction. As said, this can be achieved by local complementations as illustrated in Fig.~\ref{cluster1d}(b). For the sake of simplicity,
let us momentarily assume that the measurements are performed only at the end of the process. Then, after the interaction between $m$ and
the first two registers, $r_1$ and $r_2$, the graph at the left of the figure is created. Before the interaction with the next register
$r_3$, local complementations at $m$ and at $r_2$ take the system to the graph state at the right of the figure. The process is then
repeated with $r_3$ and iterated throughout, such that a linear cluster with $m$ always at the rightmost end is produced. In this way,
the ordering of measurements in the entire sequence will coincide with the flow of information, as in Fig. \ref{cluster1d}(a).

\par The last ingredient to add for our aim is the observation that the operations involving
any flying register commutes with the ones involving the other flying registers. This
implies that the reasoning above works equally well when measurements are not performed only at the end. In
particular, it leads us to the sequential implementation through the following cycle, schematized in Fig.
\ref{cluster1d}(c):
$(i)$ $\hat{CZ}_{mr_{i}}$ is applied to $m$ and $r_i$. $(ii)$ By local complementations at $r_i$
and $m$ their positions in the graph are swapped.  Finally $(iii)$ $r_i$ is measured, leaving $m$ with the quantum
information processed exactly as in the conventional 1WQC. The process then starts again with the next flying
register to pass by.

\par
\begin{figure}[t]
\begin{center}
\includegraphics[width=0.7\linewidth]{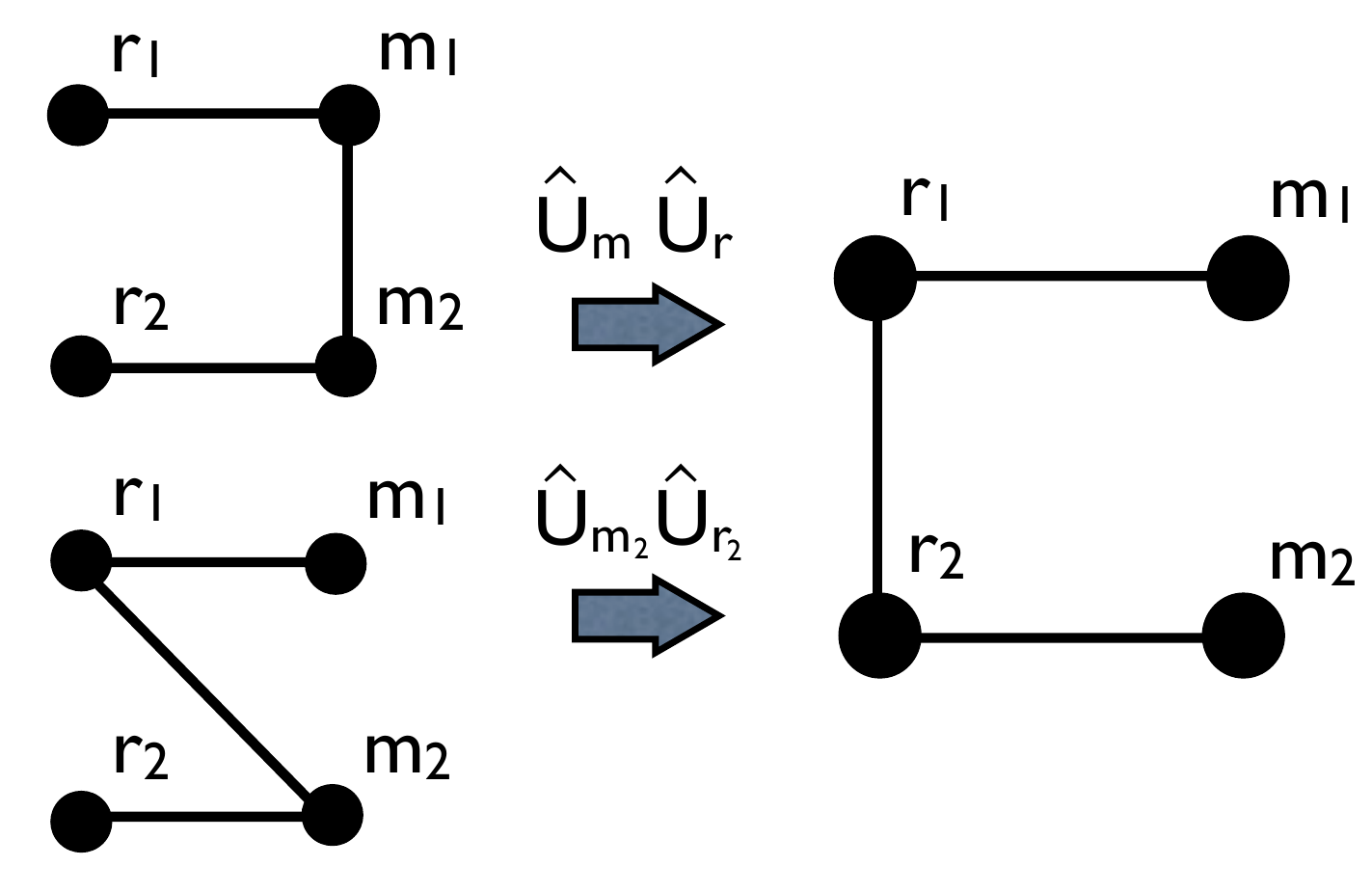}
\caption{Sequential MBQC in 2D can be carried out assuming that it is possible to
build one of the graph states from the left side.
Applying a Fourier-like transform over each of the systems (top) and over $m_2$ and $b_2$ (bottom),
both graph states transform to the one at the right.
Then, the computation proceeds by measuring $r_1$ and $r_2$.  In the first case there is a direct interaction between
the quantum memories, while in the second one the register, $r_1$, acts as a quantum bus.
\label{cluster2d}}
\end{center}
\end{figure}

 Universality in 1WQC is achieved with 2D cluster states as a resource, which are usually represented by square-lattice cluster states.
In such a case, horizontal connections corresponds to quantum wires, while vertical ones to two-body operations \cite{1WQC}.
In actual applications, however, a fully connected square lattice is not required and unnecessary nodes are removed by proper measurements  \cite{1WQC}.
Therefore, one can see that the basic building block required for sequential MBQC in 2D is
represented by the graph state  at the right of Fig. \ref{cluster2d}. As shown in the figure,
each of the graph states from the left hand side is related by a local operation
(as we have shown one can choose the Fourier transform) to the graph state at the right.
Hence, the sequence is similar to the one outlined for the quantum wire: $(i)$ generate
one of the graph states at the left hand side of Fig. \ref{cluster2d}, $(ii)$ apply the local Fourier transformations,
$(iii)$ implement the local measurements to the registers. Combining these steps with the
sequential implementation of quantum wires described above, universal quantum computation with $n$ logical quantum inputs can
be achieved sequentially using $n$ quantum memories and $n$ quantum registers that are locally measured in each instance of the computation.
As already mentioned, the registers do not interact with each other, but they can act as quantum buses mediating the interaction
between the memories (as it is the case in the bottom-left hand side graph of Fig. \ref{cluster2d}). 

To end up with, the method extends to a three-dimensional-cluster computation implemented with 
a bidimensional cluster of memories. There,  since the measurements are done sequentially layer by layer, 
the topological fault-tolerant techniques \cite{FaultTolerant} for 1WQC can be directly applied in the qubit case.

\section{Implementations}
\label{Imp}
Several physical setups lie naturally within the range of applicability of the ideas presented. For example,
cavity quantum-electrodynamics (QED) atom-photon interfaces \cite{CQED,Wilk} are perfectly-suited for
our scheme in its qubit version: Zero- and one-photon states of  fields inside microwave resonators can
provide the memories, and atoms crossing through can play the role of the flying registers \cite{CQED,Solano05}.  In a dual way, ground states
of static atoms inside optical cavities can be used to encode the memories, and single photons emitted from them the flying
registers  \cite{Wilk,Solano05,Leandro}. Also artificial-atom-photon interfaces can be used analogously,
as for instance quantum-dot \cite{Lindner09} and circuit-QED \cite{Circuit} architectures. Collective
atom-light interfaces \cite{Polzik} in turn give yet a further prominent example: Atomic ensembles in squeezed spin-coherent
states of long-lived ground states are sequentially shined through by squeezed-light pulses  \cite{Polzik}, making thus an excellent
candidate for implementations in the CV regime.

\section{Conclusions} \label{Sec-Conclusions}
\par We have introduced a systematic procedure that implements universal MBQC sequentially.
The resulting model is less demanding from an experimental point of view and maintains
the main advantages of the conventional approach: on-line
adaptiveness of entangling operations is unnecessary and only
local measurements must be adapted along the computation. The
scheme relies mainly on the sequential preparation and
manipulation of a cluster state on an $n$-dimensional lattice
that, by iterative interactions with auxiliary quantum systems via
a fixed entangling operation, can perform MBQC corresponding to
$(n+1)$-dimensional lattices. We expect that the present proposal
contributes to the experimental quest for quantum computing and
establishes new useful grounds for the application of quantum
interfaces in  general.

%%%%%%%%%%%%%%%%%%%%%%%%%%%%%%%%%%%%%%%%%%%%%%%%%%%%%%%%
%%%%%%%%%%%%%%%%%%%%%%%%%%%%%%%%%%%%%%%%%%%%%%%%%%%%%%%%
\begin{acknowledgments} 
\par We thank J. Anders and D. E. Browne for useful comments.
This work was supported by the European COMPAS project and the ERC Starting
grant PERCENT, the Spanish MEC FIS2007-60182, Consolider-Ingenio QOIT and Juan de la Cierva projects, the
Generalitat de Catalunya and Caixa Manresa.
\end{acknowledgments} 
%%%%%%%%%%%%%%%%%%%%%%%%%%%%%%%%%%%%%%%%%%%%%%%%%%%%%%%%
%%%%%%%%%%%%%%%%%%%%%%%%%%%%%%%%%%%%%%%%%%%%%%%%%%%%%%%%


\begin{thebibliography}{20}
\bibitem{comment}
We use the term ``quantum memories" to refer to quantum systems with robust coherence properties.
\end{thebibliography}
\end{document}